\documentclass[%
reprint,%
 amssymb, amsmath,%
 aip,cha,%
]{revtex4-1}

\usepackage{docs}%
\usepackage{bm}%
\usepackage[colorlinks=true,linkcolor=blue]{hyperref}%
\usepackage{graphicx}
\expandafter\ifx\csname package@font\endcsname\relax\else
 \expandafter\expandafter
 \expandafter\usepackage
 \expandafter\expandafter
 \expandafter{\csname package@font\endcsname}%
\fi
\begin{document}
\title{Synchronization in a network of delay coupled maps with stochastically switching topologies }
\author{Mayurakshi Nag and Swarup Poria}
\email{swarup$_$p@yahoo.com}
\affiliation{Department of Applied Mathematics, University of Calcutta, 92, A.P.C Road, Kolkata-700009, India.}

\begin{abstract}{The synchronization behavior of delay coupled chaotic smooth unimodal maps over a ring network with stochastic switching of links at every time step is reported in this paper. It is observed that spatiotemporal synchronization never appears for nearest neighbor connections; however, stochastic switching of connections with homogeneous delay $(\tau)$ is capable of synchronizing the network to homogeneous steady state  or periodic orbit or synchronized chaotically oscillating state depending on the delay parameter, stochasticity parameter and map parameters.  Linear stability analysis of the synchronized state is done analytically for unit delay and the value of the critical coupling strength, at which the onset of synchronization occurs is determined analytically.  The  logistic map $rx(1-x)$ (a smooth unimodal map) is chosen  for numerical simulation purpose. Synchronized steady state or synchronized  period-2 orbit is stabilized for delay $\tau=1$. On the other hand for delay $\tau=2$ the network is stabilized to the fixed point of the local map. Numerical simulation results are in good agreement with the analytically obtained linear stability analysis results. Another interesting observation is the existence of synchronized chaos in the network for delay $\tau>2$. Calculating synchronization error and plotting time series data and Poincare first return map  the existence of synchronized chaos is confirmed. The results hold good for other smooth unimodal maps also. }
\end{abstract}

\pacs{05.45.-a} \maketitle

\begin{quotation}
Due to the nonzero propagation speed of the transmitted signals the time delay in coupling arises in any system. Delay systems are in general infinite dimensional and  can display complex dynamics and a very little is known  about the basic relations between network structure and delay dynamics.
Delay induced synchronization in coupled map lattice (CML) has already been discussed in many papers. However, in  many  systems (for example, communication, ecological, social and contact networks)  links  are  not  always  active  and  the  connectivity between the units changes (stochastically or deterministically) in time with a rate ranging from slow  to  fast. These facts motivates us to find out the effects of time  delay in a CML having stochastically switching network topology.  In a stochastically switching network connection between nodes changes randomly with time.   Conditions for the stability of the synchronized steady state and periodic orbit  are derived analytically using the properties of block circulant matrix. It is shown that in  certain range of coupling (depending on probability of random connection) the CML synchronizes to periodic or chaotic orbit depending on delay time($\tau$) and map parameters. In case of  $\tau=1$ synchronized $period-2$ is observed whereas for $\tau=2$ synchronized steady state and for higher values of $\tau$ synchronized chaos are observed.

\end{quotation}

 \section{\bf{Introduction}}
~~~There has been an increasing interest in the study of spatially extended systems with local and nonlocal interactions to understand almost all complex behaviors in nature. The analysis of  collective behavior of time-delay coupled networks  is a topic of great interest  for its fundamental significance from a dynamical systems point of view and for its practical relevance in modeling various physical, biological, and engineering systems, such as coupled laser arrays, gene regulatory networks and complex ecosystems\cite{laser,aihara,eco}.
Realistic modeling  of  many  large  networks  with  nonlocal  interaction  inevitably  requires connection delays to be taken into account, since they naturally arise as a consequence of finite information transmission and processing speeds among  the units. Delay systems are in general infinite dimensional and  can display complex
dynamics and a very little is known yet about the basic relations between network structure and delay dynamics. The  first  systematic  investigation  of delay coupled   phase  oscillators    was  done  by  Schuster  and  Wagner\cite{wagner}. Delay can give rise to some significant phenomena like synchronization\cite{ouchi, amritkar,tang}, multistability\cite{shrimali,hens} and amplitude death \cite{reddy,prasad}. Dissipative systems with time delayed feedback can generate high dimensional chaos and the dimension of chaotic attractor can be made larger by increasing delay time \cite{alher,fischer,klein,mandel}. In neural network and communication network time delays ubiquitously exist \cite{pyragas,zhang}. Networks with time delay are also very useful for modeling  coordinated brain activity \cite{deco}, pattern recognitions etc.
 Delay can enhance the coherence of chaotic motion\cite{goldbin}, as well as it can stabilize synchronous state in oscillator network\cite{strogatz}. Some new phenomena like oscillation death, stabilizing periodic orbits, enhancement or suppression of synchronization, chimera state etc arise due to the time delay\cite{heil,scholl,dhamla,lakhsmanan,sen,sethia}. So there are strong reasons to shed light on the issue of such systems.

Synchronization of coupled map lattice  under delay coupling in globally coupled logistic maps\cite{kaneko,marti2}, coupled chaotic maps with inter-neural communication\cite{atay} or intra-neural communication\cite{cheng} has been intensively investigated. In modeling many realistic systems of biological, technological and physical significance using CML with either nearest neighbor interaction or global coupling  is unable to capture all essential features of the dynamics of the extended systems\cite{watt}. However,  in  many  systems (for example, communication, ecological, social and contact networks)  links  are  not  always  active  and  the  connectivity between the units changes (stochastically or deterministically) in time with a rate ranging from slow  to  fast \cite{holm}. One can think that time variations of links represent the evolution of interactions over time of the system.  Such time-varying interactions are commonly found in social networks, communication, biological systems, spread of epidemics, computer networks, world wide web etc. and  have  been  shown  to  result  in  significantly different emergent phenomena \cite{kohar}.   \\

The synchronization behavior of  time delay coupled network with dynamic random updating of links at every time  are not yet investigated. In this work, we have considered homogeneous delay coupled ring network with stochastic updating of links at every time and investigated the effects of dynamic random updating of links and delay parameter on the  synchronization phenomena of the network. Most interestingly, we study the stability of the synchronized state when the underlying  network evolves in time taking delay parameter as one unit of time.
 We  analyze synchronization behavior of dynamic random network with identical local maps. We consider
Watts-Strogatz (WS) networks and vary the fraction of random links $p$ to  cover  a  broad  range  of  networks  varying  from  a regular ring topology for $p=0$ to completely random networks for $p=1$ and for intermediate values of $p$, such networks are characterized by small-world networks : small path length and high clustering coefficient.

The paper is organized as follows: In section \ref{sec:lev1} the model is discussed. Section \ref{sec:lev2} contains linear stability analysis results for complete synchronization. Numerical simulation results are also presented and analyzed in this section.    Finally, conclusion is drawn in section \ref{sec:lev3}.


\section{\label{sec:lev1}\bf{Model}}

Governing equations for delay coupled maps with nearest neighbor interaction over a ring network are the following
\begin{equation}
x_{t+1}(i)=(1-\epsilon)f(x_t(i))+\frac{\epsilon}{2}(x_{t-\tau}(i-1)+x_{t-\tau}(i+1))
\end{equation}
where, $x_t(i)$ represents the state variable, $t(\geq0)$ the integer valued time index, $i=1,...N$ the space index, $N$ being the linear size of the array. $\epsilon$ is the coupling strength ,$\tau\geq0$ represents the constant delay time.
In our model  every  link  in  the  network  rewires  stochastically  and
independently  in every time step. In particular, at every time update  we connect a fraction $p$ of randomly chosen sites in the
lattice to two other random sites, instead of their nearest neighbors. Then the evolution rule for $p$ fraction of nodes of the network are followed by
\begin{equation}
x_{t+1}(i)=(1-\epsilon)f(x_t(i))+\frac{\epsilon}{2}(x_{t-\tau}(\xi)+x_{t-\tau}(\eta))
\end{equation}
where $\xi$ and $\eta$ are random integers uniformly distributed
in the set $\{1, 2, 3, . . . , N \}$. At the same time the dynamical equation of the time evolution rule for remaining $(1-p)$ fraction of nodes are governed by
\begin{equation}
x_{t+1}(i)=(1-\epsilon)f(x_t(i))+\frac{\epsilon}{2}(x_{t-\tau}(i-1)+x_{t-\tau}(i+1))
\end{equation}
Thus $p = 0$ corresponds to the usual nearest-neighbor
diffusion for all nodes, whereas $p = 1$ is related to completely random diffusion.
For our present study we choose one dimensional unimodal smooth map.

\section{\label{sec:lev2}\bf {Results}}
In this section, we shall do linear stability analysis of synchronized spatiotemporal fixed point of the dynamic random network of delay coupled
maps first.
\subsection{\bf{Stability Analysis}}
We consider the case with delay parameter  $\tau=1$ and check the stability of the synchronized spatiotemporal  fixed point of the map.
Stability analysis of synchronized spatiotemporal fixed point for dynamic random network is relatively complicated since it will involve diagonalization of random matrices. To calculate the stability of the synchronized fixed point, we will construct
an average probabilistic evolution rule for the sites, which
becomes a sort of mean field version of the dynamics \cite{sinha2002}. We assume that the total contribution due to $k$ randomly chosen neighbors is $k<x(t)>.$ In our case, $k=2.$ Now,
the averaged out evolution equation for any site $i$ is the following
\begin{eqnarray}
x_{t+1}(i)=(1-\epsilon)f(x_t(i))+(1-p)\frac{\epsilon}{2}(x_{t-\tau}(i+1) \nonumber\\
+x_{t-\tau}(i-1))+\frac{p\epsilon}{N}\sum_{j=1}^N x_{t-\tau}(j).
\end{eqnarray}
The delayed system (4) can now be written in following form by introducing  a new variable $y_t(i)$ as
\begin{eqnarray}\label{ab}
x_{t+1}(i) & = &(1-\epsilon)f(x_t(i))+(1-p)\frac{\epsilon}{2}\{y_t(i+1) \nonumber \\
& + & y_t(i-1)\}+\frac{p\epsilon}{N}\sum_{j=1}^N{y_t(j)} \nonumber\\
y_{t+1}(i) & = & x_t(i)
\end{eqnarray}
Now it becomes an equation with no delay term. In this way every coupled map lattice model with homogeneous delay  can be easily transformed into an CML with no delay.
Synchronization occurs when the state variables of each node adopt the same value
for all the coupled maps at all times $t$, i.e. $x_t(1) = x_t(2) =x_t(3) = .... = x_t(N)=x^*$ and $y_t(1) = y_t(2) =y_t(3) = ... = y_t(N)$ for all times greater than a transient time. A range of the coupling strength may exist for which such a
synchronized steady state can be obtained depending on the nature of the individual map and an adequate rewiring probability $p$ and delay parameter $\tau$. The linear stability analysis of  synchronized spatiotemporal steady state of the model (5)   is performed for smooth unimodal maps.
The Jacobian matrix for the system (\ref{ab}) can be cast as
\begin{equation}
\label{jac}
J=\left[
\begin{array}{cccccc}
A+C & B+C & C & C &...& B+C\\
B+C & A+C & B+C & C &...& C\\
\vdots & \vdots &  \vdots & \vdots & ... &\vdots \\
B+C & C & C & C &...& A+C
\end{array}
\right]
\end{equation}
where,

$A=\left(
\begin{array}{cc}
(1-\epsilon)a & 0\\
1 & 0
\end{array}
\right)$,
$B=\left(
\begin{array}{cc}
0 & (1-p)\epsilon/2 \\
0 & 0
\end{array}
\right)$ and\\
$C=\left(
\begin{array}{cc}
0 & p\epsilon /N\\
0 & 0
\end{array}
\right)$,
$a=\frac{df(x)}{dx}|_{x=x^*}$.\\
To derive the stability condition of the synchronized steady state we will study the eigenvalues of the matrix $J$.
The matrix $J$ is a block circulant matrix. It can be reduced to a block diagonal matrix using unitary transformation.
The block diagonal form of $J$ is the following
\begin{equation}
D=\left(
\begin{array}{cccc}
M_0 & 0 & ...& 0\\
0 & M_1 & ...& 0\\
\vdots & \vdots  & ... &\vdots \\
0 & 0  &...& M_{N-1}
\end{array}
\right)
\end{equation}
where, the matrix $M_r(r=0,1,2,...N-1)$ are $2\times 2$ matrices given by
\begin{eqnarray}
M_r &=& (A+C)+\omega_r(B+C)+\omega_r^2C+....+\omega_r^{N-1}(B+C)         \nonumber \\
&=& A+2Bcos\theta_r+C(1+e^{i\theta_r}+e^{2i\theta_r}+...+e^{(N-1)i\theta_r}) \nonumber
\end{eqnarray}
$\omega_r=e^{i\theta_r}$, $\theta_r=2\pi r/N$.\\
After substituting the values of $A, B$ and $C$ we obtain
$M_r=\left(
\begin{array}{cc}
(1-\epsilon)a & (1-p)\epsilon cos\theta_r+\frac{p\epsilon}{N}\frac{1-e^{Ni\theta_r}}{1-e^{i\theta_r}}\\
1 & 0
\end{array}
\right)$.
Clearly, the matrix
 $M_0=\left(
\begin{array}{cc}
(1-\epsilon)a & \epsilon\\
1 & 0
\end{array}
\right)$.
Now,  $\theta_r$ can vary between 0 and $2\pi$.
The eigenvalues of $M_0$ are $(a(1-\epsilon)\pm \sqrt{a^2(1-\epsilon)^2+4\epsilon})/2$. If there exist at least one eigenvalue of J  greater than 1 for all $\epsilon$ then the  synchronized steady state  is not stable for the system (\ref{ab}). This situation appears when $|a|>1$ and consequently the network has no stable synchronized steady state. On the other hand for $|a|<1$  of eigenvalues of $M_0$ lie inside the unit circle centered at origin in the complex plane . Also we calculated the eigenvalues of $M_r$ for the case $r \neq 0$ and found them lying within unit circle centered at origin for $|a|<1$. This indicates that for $|a|<1$ stable synchronized steady state  ought to occur; whereas, for $|a|>1$ synchronized fixed point can never be achieved. 
\\

Next, we seek the stability criterion of synchronized period-2 orbit of the lattice, which are the period 2  points of the local map. To do this, we have to find the eigenvalues of
the matrix $J'=J_1J_2$, where $J_1=J|_{x=x_1^*}$ and $J_2=J|_{x=x_2^*} $ and $\{x_1, x_2\}$ is a period 2 orbit of the local map.
Thus, the Jacobian turns out to be
\begin{equation}
\label{jaco}
J'=\left[
\begin{array}{cccccc}
A'+C' & B'+C' & C' & C' &...& B'+C'\\
B'+C' & A'+C' & B'+C' & C' &...& C'\\
\vdots & \vdots &  \vdots & \vdots & ... &\vdots \\
B'+C' & C' & C' & C' &...& A'+C'
\end{array}
\right]
\end{equation}

where,

$A'=\left(
\begin{array}{cc}
(1-\epsilon)^2cd & 0\\
(1-\epsilon)d & 0
\end{array}
\right)$,
$B'=\left(
\begin{array}{cc}
(1-p)\epsilon/2 & (1-\epsilon)c(1-p)\epsilon/2 \\
0 & (1-p)\epsilon/2
\end{array}
\right)$ and
$C'=\left(
\begin{array}{cc}
p\epsilon /N & (1-\epsilon)cp\epsilon /N\\
0 & p\epsilon /N
\end{array}
\right)$,
$c=\frac{df(x)}{dx}|_{x=x_1^*}$ , $d=\frac{ df(x)}{dx}|_{x=x_2^*}$.\\

The matrix $J'$ is also block circulant matrix. So approaching as above we reach at a block diagonal matrix
\begin{equation}
D'=\left(
\begin{array}{cccc}
M'_0 & 0 & ...& 0\\
0 & M'_1 & ...& 0\\
\vdots & \vdots  & ... &\vdots \\
0 & 0  &...& M'_{N-1}
\end{array}
\right)
\end{equation}
where, the matrix $M'_r (r=0,1,2,...N-1)$ are $2\times 2$ matrices given by
\begin{eqnarray}
M'_r \nonumber\\
&=& (A'+C')+\omega_r(B'+C')+\omega_r^2C'+....+\omega_r^{N-1}(B'+C') \nonumber \\
&=& A'+2B'cos\theta_r+C'(1+e^{i\theta_r}+e^{2i\theta_r}+...+e^{(N-1)i\theta_r}) \nonumber
\end{eqnarray}
where, $\omega_r=e^{i\theta_r}$, $\theta_r=2\pi r/N.$

The eigen values of the matrix\\
$M'_r=\left(
\begin{array}{cc}
(1-\epsilon)^2cd+(1-p)\epsilon cos\theta_r & (1-\epsilon)c\epsilon(1-p)cos\theta_r  \\
(1-\epsilon)d & (1-p)\epsilon cos\theta_r
\end{array}
\right)$
are
$\lambda=\frac{1}{2}\{(1-\epsilon)^2cd+2(1-p)\epsilon cos\theta_r \pm \sqrt{(1-\epsilon)^4c^2d^2+4(1-p)\epsilon cos\theta_r(1-\epsilon)^2cd}\}.$
For the stability of period-2 all the eigenvalues must lie within the unit circle of the complex plane. i.e.
$|\lambda|\leq 1$. 
 We check the eigenvalues for $cos\theta_r$ between -1 to 1.
For $cos\theta_r=-1$, the inequality gives
\begin{equation}
\epsilon \geq \frac{u-1}{u-1+p}
\end{equation}
and for $cos\theta_r=1$, we reach at the inequality
\begin{equation}
\epsilon \geq \frac{u-1}{u+1-p}.
\end{equation}
where, $u=\sqrt{|cd|}$.
Therefore, we point out that the period-2 fixed point is stable in the range
\begin{equation}
 Max\{\frac{u-1}{u-1+p},\frac{u-1}{u+1-p}\}\leq \epsilon <1
\end{equation}
which implies,
\begin{equation}
 \frac{u-1}{u-1+p}\leq \epsilon <1.
\end{equation}
The critical coupling strength depends on the properties of the local map and randomness of rewiring $(p)$ and the delay parameter. The stability range of synchronized period -2 increases monotonically with the randomness of rewiring $(p)$. Notice that this condition is independent of the lattice size $N$.
Specifically, we take chaotic logistic map $f(x)=4x(1-x)$ having nonzero fixed point $x_0=0.75$.
Since $a=-2$ for this point of the map, i.e $|a|>1$, period-1 fixed point never synchronizes for this chaotic logistic map.
Now, for this map $cd=-4.096$, period-2 fixed points being $x_1^*=0.9045$ and $x_2^*=0.3454$.
 We find that the period-2 fixed point is stable in the range
 \begin{equation}
 Max\{\frac{1.0238}{1.0238+p},\frac{1.0238}{3.0238-p}\}\leq \epsilon <1.
\end{equation}
which implies,
\begin{equation}
 \frac{1.0238}{1.0238+p}\leq \epsilon <1.
\end{equation}
For fully random coupling the range of stability of synchronized period-2 orbit becomes $0.5058<\epsilon<1$.

\subsection{\bf{Simulation Results and Analysis}}
\begin{figure}[ht]
             \includegraphics[width=3.0in,height=2.0in]{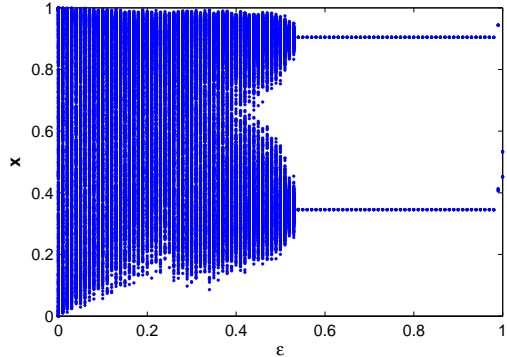}
             \caption{Bifurcation diagram of delay coupled logistic maps $4x(1-x)$ for $\tau=1$ and $p=1$.}
             \label{fg1}
             \end{figure}

\begin{figure}[ht]
             \includegraphics[width=3.0in,height=2.0in]{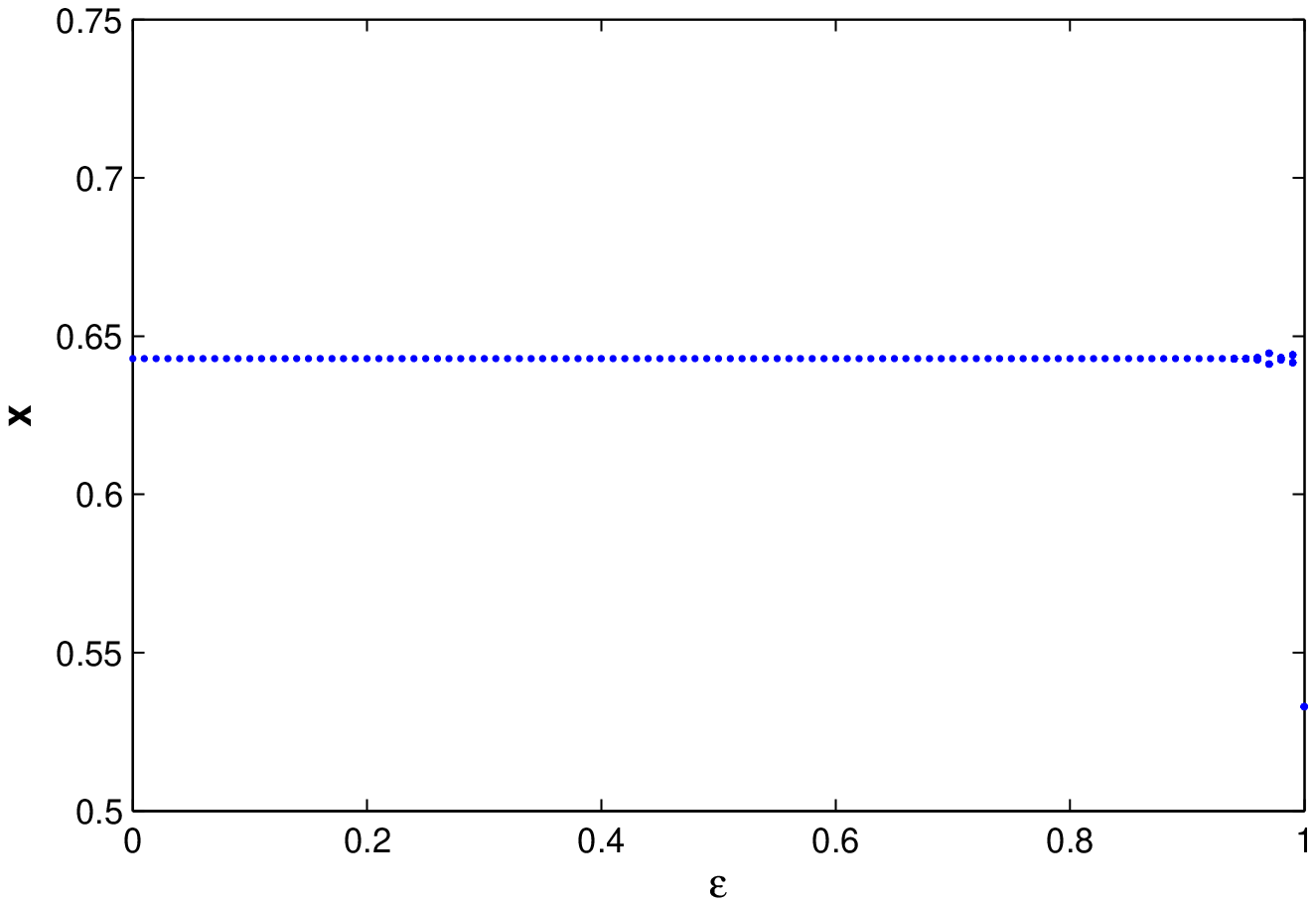}
             \caption{Bifurcation diagram of delay coupled logistic maps $2.8x(1-x)$ for $\tau=1$ and $p=1$.}
             \label{fg1b}
             \end{figure}

\begin{figure}[ht]
             \includegraphics[width=3.0in,height=2.0in]{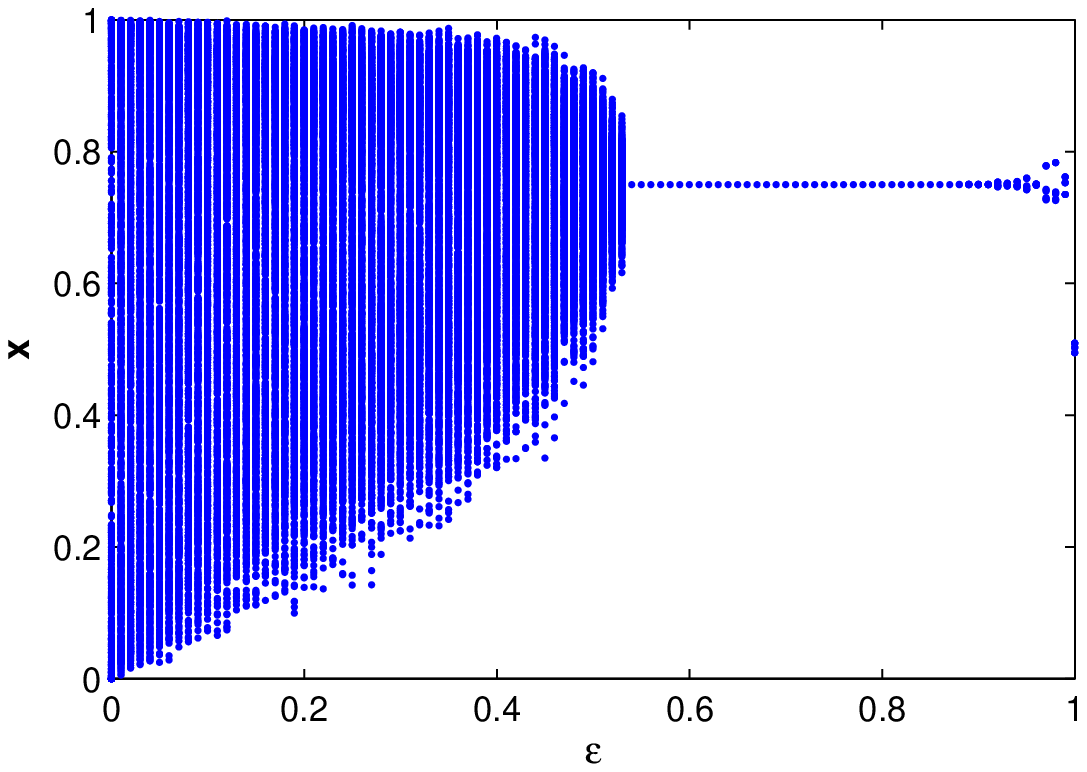}
             \caption{Bifurcation diagram of delay coupled logistic maps for $\tau=2$ and $p=1$.}
             \label{fg2}
             \end{figure}

\begin{figure}[ht]
             \includegraphics[width=3.0in,height=2.0in]{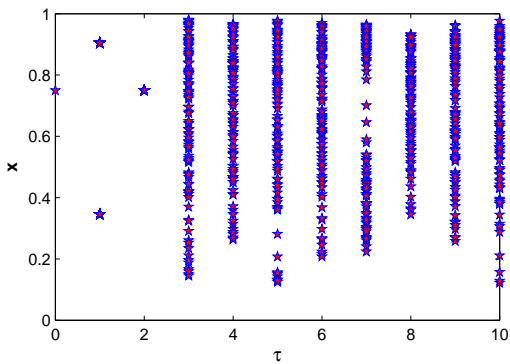}
             \caption{Bifurcation diagram of delay coupled logistic map with variation in delay time $\tau$ with $p=1$.}
             \label{fg3}
             \end{figure}

 For numerical simulation we have chosen the  logistic map $f(x)=rx(1-x)$ and the lattice size to be 100. First we take the delay time to be $\tau=1$. Bifurcation diagram of the state variables with the variation of the coupling strength $\epsilon$ of the lattice for fully random rewiring is drawn in figure \ref{fg1} for $r=4$ and in figure \ref{fg1b} $r=2.8$. We observe synchronized fixed point for $r=2.8$ but synchronized period-2 behavior  for a long range of coupling strength for $p=1$ which is consistent with our analytical findings. From numerical simulation the range of synchronized period-2 orbit is found to be $0.535<\epsilon<1$ which is very close to the analytically estimated range. Next, we draw the bifurcation diagram of the state variables for $\tau=2$. Transition to synchronized fixed point for complete random rewiring is depicted in figure \ref{fg2}. The nature of bifurcation is the same as the bifurcation of a system without delay. The bifurcation diagram with the variation of delay time $\tau$  for $\epsilon=0.6$ and $p=1$ is shown in figure \ref{fg3}. This figure exhibits the existence of synchronized chaos for $\tau>2$.

\begin{figure}[ht]
             \includegraphics[width=3.0in,height=2.0in]{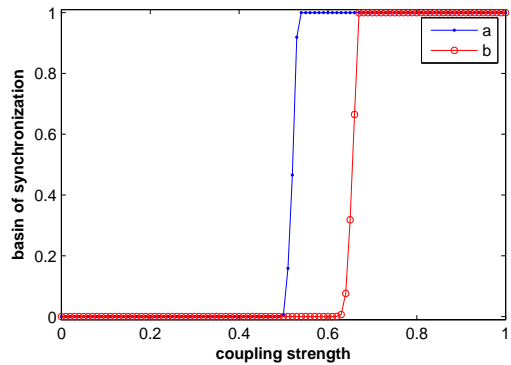}
             \caption{Sizes of the basin of attraction of the synchronized period 2 state with the variation of coupling strength for (a) $p=1.0$ and (b) $p=0.5$.}
             \label{fg4}
             \end{figure}

\begin{figure}[ht]
             \includegraphics[width=3.0in,height=2.0in]{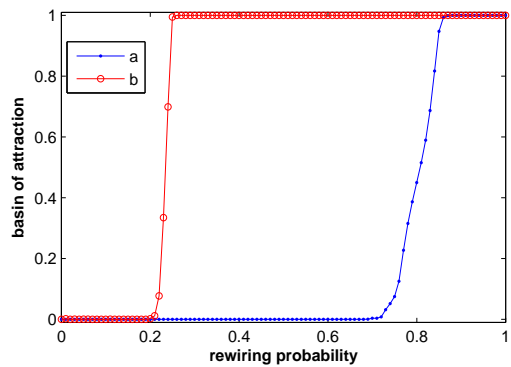}
             \caption{Sizes of the basin of attraction of the synchronized period 2 state with the variation of rewiring probability for (a) $\epsilon=0.55$ and (b) $\epsilon=0.8$ with $p=1$.}
             \label{fg5}
             \end{figure}

Furthermore, we calculate the basin size of synchronization. If out of $s$ random initial conditions synchronization is obtained in $k$ cases then basin size will be $\frac{k}{s}$, i.e the basin size lies between the values 0 and 1.
To compute basin size we have taken 100 random initial conditions. Variation of basin sizes with respect to the coupling strength for different connectivity $p=1$ and $p=0.5$ are displayed in figure \ref{fg4}. The range of synchronization increases with the increase in $p$. For different coupling strength $\epsilon=0.55$ and $\epsilon=0.8$ we plot the basin sizes with the variation of the rewiring probability in figure \ref{fg5}.

\begin{figure}[ht]
             \includegraphics[width=3.0in,height=2.0in]{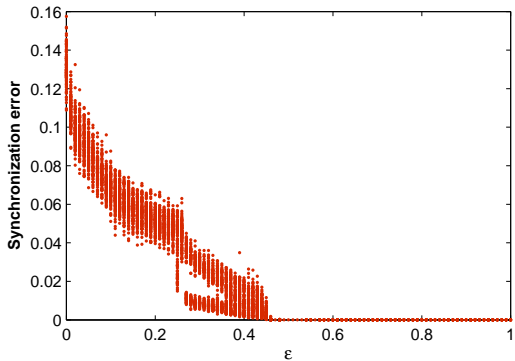}
             \caption{Synchronization error of the lattice with respect to coupling strength for $p=1$ and $\tau=5$.}
             \label{fg6}
             \end{figure}

\begin{figure}[ht]
             \includegraphics[width=3.0in,height=2.0in]{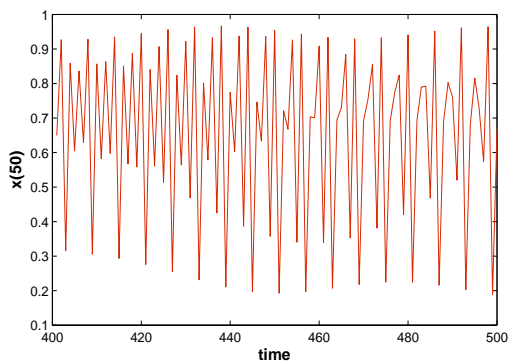}
             \caption{Time evolution of one node of the lattice for $p=1$, $\tau=5$ and $\epsilon=0.8$.}
             \label{fg7}
             \end{figure}

\begin{figure}[ht]
             \includegraphics[width=3.0in,height=2.0in]{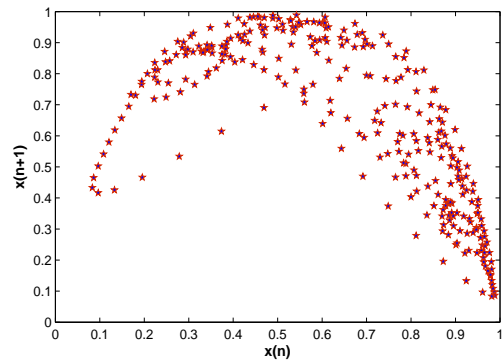}
             \caption{Poincare section of the coupled map for $p=1$, $\tau=5$ and $\epsilon=0.8$.}
             \label{fg8}
             \end{figure}

Coming to the scenario of synchronized chaos we have to justify the occurrence of chaos as well as the synchronization among the lattice points.
For the confirmation of existence of synchronized chaos for $\tau>2$ we plot the synchronization error with the variation of coupling strength. In figure \ref{fg6} we have displayed the nature of synchronization error for $p=1$ and $\tau=5$. This shows that for high coupling strength fully synchronized state is achieved irrespective of the nature of the orbit. We have plot the time series of one site for $p=1$, $\tau=5$ and $\epsilon=0.8$ in figure \ref{fg7} to show that chaos is generated at each node for higher delay time. We confirm the existence of chaos by plotting the Poincare first return map in figure \ref{fg8}.

\section{\label{sec:lev3}\bf {Conclusion}}

We have reported the synchronization behavior of homogeneous time delay coupled one dimensional ring network of  unimodal maps in presence of stochastically switching of links. We have considered homogeneous delay coupled ring network with stochastic updating of links at every time and investigated the effects of dynamic random updating of links and delay parameter and coupling strength on the  synchronization phenomena of the network. The key finding is the analytically derivation of the stability range of the synchronized  steady state and synchronize period 2 orbit of the delay coupled CML for $\tau=1$. It is observed that the critical coupling strength for obtaining synchronized period-2 depends on the properties of the local map and randomness of rewiring $(p)$ and the delay parameter. The stability range of synchronized period -2 increases monotonically with the randomness of rewiring $(p)$ but this condition is independent of the lattice size $N$. The analytically calculated stability range of synchronized period-2 orbit  is in good agreement  with the numerical one.   We have taken logistic map for numerical simulations.   The random rewiring of spatial connection has significant effect on spatiotemporal synchronization as usual. Either synchronized steady state or synchronized period 2 orbit is observed for $\tau=1$ depending on the map parameters.  Synchronized period-1 orbit is observed for $\tau=2$. Existence of synchronized chaos is reported for $\tau>2$. These   results also hold  for other smooth unimodal maps. We calculated the synchronization error to show the spatial synchronization of the nodes. The time evolution of one node is shown as an evidence of the existence of chaos. Further analysis using  Poincare first return map confirms chaos. Thus we confirm the existence of synchronized chaos for $\tau>2$ for higher coupling strength. The coupling topology can crucially 
influence the synchronizability of the CML. In contrast to the non-delayed CML with stochastic updating of links   where only synchronized steady state exists (at strong coupling strength) \cite{sinha2002} here, because of homogeneous time delay  the CML supports wide range of qualitatively different synchronized states including synchronized chaos. 
The results presented in this paper may help us to understand the origin of diversity of synchronous dynamical states observed in large number of biological, chemical and social networks, which can emerge either due to finite information transmission delays or small-world topology.

\end{document}